\documentclass[conference]{IEEEtran}
\IEEEoverridecommandlockouts
\usepackage{cite}
\usepackage{url}
\usepackage{amsmath,amssymb,amsfonts}
\usepackage{algorithmic}
\usepackage{graphicx}
\usepackage{textcomp}
\usepackage{xcolor}
\usepackage[linesnumbered,ruled,vlined]{algorithm2e}
\def\BibTeX{{\rm B\kern-.05em{\sc i\kern-.025em b}\kern-.08em
    T\kern-.1667em\lower.7ex\hbox{E}\kern-.125emX}}
\begin{document}

\title{Comprehensive Survey on Adversarial Examples in Cybersecurity: Impacts, Challenges, and Mitigation Strategies\\
}

\author{\IEEEauthorblockN{Li Li}
\textit{Virginia Tech}\\
}

\maketitle

\begin{abstract}
Deep learning (DL) has significantly transformed cybersecurity, enabling advancements in malware detection, botnet identification, intrusion detection, user authentication, and encrypted traffic analysis. However, the rise of adversarial examples (AE) poses a critical challenge to the robustness and reliability of DL-based systems. These subtle, crafted perturbations can deceive models, leading to severe consequences like misclassification and system vulnerabilities. This paper provides a comprehensive review of the impact of AE attacks on key cybersecurity applications, highlighting both their theoretical and practical implications. We systematically examine the methods used to generate adversarial examples, their specific effects across various domains, and the inherent trade-offs attackers face between efficacy and resource efficiency. Additionally, we explore recent advancements in defense mechanisms, including gradient masking, adversarial training, and detection techniques, evaluating their potential to enhance model resilience. By summarizing cutting-edge research, this study aims to bridge the gap between adversarial research and practical security applications, offering insights to fortify the adoption of DL solutions in cybersecurity.

\end{abstract}


\section{Introduction}
With the continuous maturation of deep learning technology, it has played a significant role in various fields. The ongoing development of cybersecurity has also increased its demand for deep learning technology, which enables and facilitates many security-based applications\cite{li,wu}. 
In detail, deep learning technology can first better \textbf{classify malware} rapidly and accurately with their deeper architectures, outperforming human analysts. 
Second, the current deep-learning techniques utilized to \textbf{detect the Domain Generation Algorithms (DGAs) of the botnets} can learn features automatically, thereby downplaying human participation in feature engineering \cite{curtin2019detecting,kim2020botnet,tran2018lstm,vinayakumar2020visualized,yang2020detecting}.
Third, DL techniques can be used in \textbf{network intrusion detection} \cite{zhang}, in handling heterogeneous and nonlinear data, autonomously discovering correlations, and reducing network traffic complexity by performing both feature extraction and classification, while traditional learning techniques' shallow architecture struggles with the vast multidimensional data from new network technologies \cite{29709532,2jiang2022fgmd,2qiu2020adversarial,2shu2020generative}.
Fourth, Deep Learning (DL) is crucial in \textbf{user identification and authentication} \cite{3george2019biometric,3menotti2015deep,3park2019presentation,3souza2017deep}, with various studies showcasing remarkable performance. Notably, CNN and its variants are widely used due to their suitability for processing data from diverse arrays like images, enabling the derivation of more reliable feature abstractions for user identification/authentication.
Fifth, DL techniques are well-suited for \textbf{encrypted traffic classification and analysis} with their capability for automatic feature extraction, whose architectures based on CNNs and RNNs have demonstrated significant promise in these applications, efficiently extracting underlying spatial and temporal correlations \cite{4aceto2018mobile,4rezaei2020multitask,4shapira2019flowpic,4sirinam2018deep}. 
In summary, DL's capacity to accurately model complex data, such as time series, has led to their widespread adoption in various cyber defense applications like malware detection, botnets detection, and network intrusion detection. However, the efficacy of a detection system hinges significantly on its resilience to attacks directed at the detection system itself.

Adversarial Examples (AE) attacks are aimed at disrupting the performance of deep learning (DL) models. Such attacks involve subtle and carefully crafted modifications to input data, with the goal of deceiving DL models to generate incorrect outputs. The emergence of adversarial examples exposes the vulnerability of DL models, especially when confronted with inputs designed with malicious intent.
Adversarial Examples attacks always generate ideal interference effects at the theoretical level, significantly reducing the performance of deep learning models. However, in practical applications, whether the destructive power of disturbance attacks is sufficient is a question worthy of exploration. This article focuses on summarizing the impact of Adversarial Examples attacks on the performance of deep learning models in the field of cybersecurity, aiming to investigate whether disturbance attacks have sufficient destructive power.
In the realm of cybersecurity applications, AE attacks can lead to false positives or false negatives, thereby undermining the reliability of DL models in detecting malicious behavior. To enhance the resilience of DL models against such attacks, researchers continuously work on developing adversarial training techniques and robustness enhancement methods to improve the model's ability to discern potential threats, ensuring its effective operation when faced with intricate and malicious inputs.

Paper structure: In Chapter 2, we will introduce the impact of Adversarial Examples (AE) attacks on malware detection. In Chapter 3, we will consider the influence of AE attacks on detecting zombie networks. The content of Chapter 4 will explore the effects of AE attacks on intrusion detection. Chapter 5 will discuss the impact of AE attacks on user authentication. In Chapter 6, we will introduce the effects of attacks on encrypted data analysis. In Chapter 7, potential solutions will be provided.  In Chapter 8, we will conclude.

\section{Adversarial examples attacks in malware detection and analysis}

The models exhibit susceptibility to adversarial examples \cite{5demetrio2021adversarial}, and carefully crafted inputs designed to provoke misclassifications. Recent studies have demonstrated how attackers can meticulously manipulate malware samples to evade learning-based detection and achieve malicious objectives \cite{5abusnaina2019adversarial,5abusnaina2021dl,5alasmary2020soteria,5grosse2016adversarial,5ji2019securing,5khormali2019copycat}. Nonetheless, manipulating malware while preserving its intended functionality poses a significant challenge. Each perturbation carries the risk of introducing irreversible changes in the underlying syntax and semantics, thereby compromising the intended operation of the binary. In response to this challenge, attackers typically adopt two strategies: employing dynamic analysis methods like emulation and noninvasive perturbations to ensure the binary's functionality remains unaffected, or confining perturbations to non-functional segments of the file, such as trailing bytes. Naturally, this creates a trade-off between robust yet intricate attacks on one end and weaker but resource-efficient attacks on the other.

In \cite{maldet_1}, the authors design a methodological framework focused on evaluating the vulnerability of neural networks to adversarial attacks in the context of malware classification. They start by developing a neural network-based classifier, trained on the DREBIN dataset, a comprehensive collection of Android applications including a significant proportion of malware samples. The features for the classifier are derived through static analysis, resulting in binary indicators that represent the presence or absence of various application characteristics. The core of their method involves the creation of adversarial samples using an algorithm based on the forward derivative of the neural network, which strategically adds features to applications without compromising their functionality. This approach allows for assessing the classifier's vulnerability by measuring the misclassification rate of malware samples. Additionally, the study incorporates an examination of defensive mechanisms such as feature reduction, distillation, and re-training with adversarially crafted samples, aiming to enhance the classifier's resistance to these adversarial perturbations.The authors found that their neural network-based malware classifier, when subjected to adversarially crafted inputs, exhibited a significant vulnerability. Specifically, they achieved up to an 85\% misclassification rate for malicious applications, demonstrating the effectiveness of their adversarial crafting approach in the security-critical domain of malware detection. The experiment also revealed that the classifier's performance varied with different configurations; for instance, a setup with two layers and 200 neurons each achieved around 98\% accuracy, with false negatives at about 7\% and false positives at 3.3\%. In exploring defensive strategies, they observed that while feature reduction did not significantly enhance network resilience, techniques such as distillation and re-training on adversarially crafted samples showed more promise in reducing the network's vulnerability to adversarial manipulations. These results underscore the potential impact of adversarial attacks on neural network-based malware detection systems and the importance of incorporating robust defensive measures.

In \cite{maldet_2}, the authors designed an experimental framework to assess the vulnerability of neural network-based malware classifiers to adversarial examples. They began by training a neural network on the DREBIN dataset, which contains a wide range of Android applications, including a significant number of malware samples. The classifier was trained using static features extracted from the applications, represented as binary vectors. The adversarial examples were crafted using a modified algorithm that works in discrete input spaces, ensuring that the malware functionality remains intact. The effectiveness of these adversarial examples was tested against the classifier, achieving a substantial misclassification rate. Additionally, the study explored the efficacy of defensive mechanisms like distillation and adversarial training to enhance the resilience of these classifiers against such adversarial manipulations. This design not only demonstrated the neural network's susceptibility to adversarial attacks but also evaluated potential strategies for mitigating such vulnerabilities. The authors presented specific data showing the effectiveness of adversarial examples against neural network-based malware classifiers. Their trained model, using the DREBIN dataset, achieved an accuracy of 98.35\% with a 9.73\% false negative rate and a 1.29\% false positive rate. However, when subjected to adversarial examples, the classifier's misclassification rate for malware samples was as high as 63\%. This rate varied with different configurations of the neural network, with misclassification rates ranging from 63\% to 69\% depending on the malware ratio used during training. The study also found that defensive distillation had a limited impact on improving the classifier's resilience, only slightly reducing the misclassification rates. In contrast, adversarial training showed more effectiveness; for instance, a network trained with a malware ratio of 0.3 and additional adversarial samples saw a reduction in misclassification rate from 73\% to 67\%. These specific findings underscore the significant vulnerability of neural network-based malware classifiers to carefully crafted adversarial examples and the varying effectiveness of different defensive strategies.

In \cite{maldet_3}, the researchers designed an innovative approach to test the robustness of visualization-based malware detection systems against adversarial examples. They developed a deep learning-based classifier using Convolutional Neural Networks (CNNs), tailored for Windows and IoT malware datasets, achieving high accuracy rates. The core of their design was the creation of adversarial examples using the COPYCAT approach, which involved two main techniques: AE padding and sample injection. AE padding added adversarial data to the end of the binary, in non-executable sections, to induce misclassification without affecting executability. Sample injection involved embedding binaries from a target class into the attacked sample to achieve the same goal. These techniques were evaluated for both targeted and untargeted misclassification attacks. The study also included a transferability aspect to assess the generalization of the method across different models. This comprehensive design aimed to challenge and enhance the robustness of malware detection systems in the face of advanced adversarial examples. In their experiment, the researchers of "COPYCAT: Practical Adversarial Attacks on Visualization-Based Malware Detection" demonstrated that their approach could significantly compromise the reliability of visualization-based malware detection systems. Using their CNN-based classifiers for Windows and IoT malware datasets, they successfully generated adversarial examples that led to high misclassification rates: 98.9\% for Windows and 96.5\% for IoT datasets. The COPYCAT method, which included AE padding and sample injection techniques, was key in preserving the executability of adversarial samples while achieving these misclassification rates. Furthermore, their transferability studies showed that these adversarial examples could generalize across different models, indicating a broader applicability and threat. Notably, all adversarial examples generated were executable, distinguishing COPYCAT from other adversarial methods. 

In \cite{maldet_4}, the authors designed a comprehensive approach to assess the robustness of IoT malware detection systems that use Control Flow Graph (CFG)-based features. They initially conducted an in-depth analysis of malware binaries to construct abstract CFG structures, analyzing multiple aspects like the number of nodes and edges, and graph algorithmic constructs. The core of their design included two methodologies for crafting adversarial IoT software samples: first, they employed eight well-known off-the-shelf adversarial learning methods to manipulate the model into misclassification. Second, they introduced the Graph Embedding and Augmentation (GEA) approach, which aimed to maintain the functionality and practicality of the adversarial samples by embedding benign samples into malicious ones. This dual-method approach allowed for a thorough evaluation of the detection system's vulnerability to different types of adversarial attacks while considering the practicality and functionality of adversarial samples. In their experiment, the authors of "Adversarial Learning Attacks on Graph-based IoT Malware Detection Systems" achieved striking results demonstrating the vulnerability of CFG-based IoT malware detection systems. Using the off-the-shelf adversarial learning methods, they managed to achieve a 100\% misclassification rate, successfully misleading the detection model in every instance. This was a significant finding, indicating a complete vulnerability of the model to these attack methods. Additionally, the Graph Embedding and Augmentation (GEA) approach, designed to maintain the functionality of the adversarial samples, also proved highly effective, achieving a complete misclassification of all malware samples as benign. These outcomes highlighted the critical vulnerability of graph-based IoT malware detection systems to adversarial learning attacks, underscoring the need for more robust and less manipulable detection methods in the realm of IoT security.

In \cite{maldet_5}, the authors designed DEEPARMOUR, a malware cognitive system, to defend against adversarial attacks in malware classification. The core design includes a weighted voting system integrating three distinct machine learning classifiers: random forest, multi-layer perceptron (MLP), and structure2vec. Each classifier contributes to a robust overall decision, leveraging varied learning strategies and features. To further enhance resilience, DEEPARMOUR employs two feature reconstruction methods: term frequency-inverse document frequency (TFIDF) weighting and graph representation, both aimed at mitigating the impact of adversarial perturbations. Additionally, the system incorporates adversarial retraining, a method where it is trained with adversarial samples to improve its defense capabilities. This multi-faceted design strategy is tested on a comprehensive malware execution trace dataset, emphasizing the system's ability to withstand adversarial evasion attacks and adapt through retraining. In the experiments conducted for the "Securing Malware Cognitive Systems against Adversarial Attacks" paper, DEEPARMOUR demonstrated robust performance against adversarial attacks. The system, tested on a malware execution trace dataset with 12,536 samples, achieved a high accuracy of 0.989 in normal conditions through 10-fold cross-validation. When subjected to adversarial attacks, the system initially showed an accuracy of 0.675. Notably, the implementation of adversarial retraining significantly improved its resilience: by retraining with only 10\% of adversarial samples, DEEPARMOUR's accuracy increased to 0.839. Further enhancement was observed when 50\% of adversarial samples were used for retraining, with the accuracy reaching 0.904. These results highlight DEEPARMOUR's effectiveness in adapting and maintaining high accuracy levels in the face of sophisticated adversarial evasion attacks.

In \cite{maldet_6}, the authors designed a sophisticated system to enhance IoT malware detection robustness against adversarial attacks. The system, DL-FHMC, incorporates a multi-faceted approach, starting with CFG (Control Flow Graph) extraction from IoT software samples. It then employs a feature extraction component to calculate 23 different algorithmic features from these CFGs. The core of DL-FHMC comprises two primary components: a malware detection module, which classifies samples into malware or benign categories, and a malware classification module, which further categorizes detected malware into specific families. To bolster its defense against adversarial attacks, DL-FHMC includes a Suspicious Behavior Detector, which uses graph mining techniques to identify potentially malicious patterns in samples classified as benign. This comprehensive design integrates various machine and deep learning models, ensuring a robust and multi-layered approach to IoT malware detection and classification. In the "DL-FHMC: Deep Learning-Based Fine-Grained Hierarchical Learning Approach for Robust Malware Classification" study, the authors achieved notable outcomes in their experimental evaluations. The DL-FHMC system was tested for its effectiveness in detecting and classifying IoT malware using a dataset comprising 7,091 IoT malware and 3,000 benign samples. The system demonstrated high accuracy rates in standard conditions, with 98.90\% for the Random Forest model, 97.42\% for the Deep Neural Networks model, and 98.31\% for the Convolutional Neural Network model. In testing against adversarial attacks using Graph Embedding and Augmentation (GEA) and Sub-Graph Embedding and Augmentation (SGEA) techniques, the system showed varying degrees of robustness. For instance, with GEA, misclassification rates reached up to 100\% for larger injected graphs, indicating susceptibility to certain adversarial manipulations. However, the system's Suspicious Behavior Detector significantly contributed to identifying potentially malicious patterns in benign-classified samples, enhancing the overall robustness of the system against sophisticated adversarial attacks.

\section{Adversarial examples in botnet detection and DGAs}

Deep learning models are vulnerable to adversarial examples intentionally designed to induce mistaken predictions. Regarding DGA classification, adversaries (botnet operators) can circumvent DGA detectors by modifying the AGD names to be less detectable. The feasibility of this idea was proven \cite{6anderson2016deepdga} who presented the DeepDGA technique, and \cite{6peck2019charbot} who introduced the CharBot method. More precisely, DeepDGA employs a GAN architecture for generating benign-appearing domain names, while CharBot resorts to changing two random characters of valid domain names.

In \cite{botnet_1}, the authors designed a novel approach using Generative Adversarial Networks (GANs) to create challenging domain names for malware detection systems. The process began with the development of DeepDGA, a deep-learning model trained to generate domain names. This model was structured initially as an auto-encoder, trained on legitimate domain names from the Alexa top one million list. The auto-encoder was then repurposed into a GAN, composed of a generator and a detector, which were trained in adversarial rounds. During these rounds, the generator learned to produce domain names that were increasingly difficult for the detector to classify correctly. This approach aimed not only to create domain names that could evade detection by the GAN’s own detector but also to test if these adversarially generated domains could bypass traditional detection systems like random forest classifiers. Additionally, the study explored the potential of using these adversarially generated domains to augment training sets, thereby enhancing the detection capability of other models against new and unseen malware families. In their experiment, the authors of "DeepDGA: Adversarially-Tuned Domain Generation and Detection" demonstrated significant results with DeepDGA, their adversarially-tuned domain generation model. Initially, the autoencoder-based DeepDGA, trained on Alexa's top 1 million domains, could produce domain names that resembled legitimate domains. When repurposed into a GAN and trained through adversarial rounds, DeepDGA was able to generate domain names that effectively evaded detection. Specifically, the random forest classifier's ability to detect these adversarially generated domains degraded with each adversarial round, with an apparent asymptote in detection reduction after three rounds. For instance, the area under the ROC curve (AUC) for the random forest classifier reduced from 0.9710 in round 0 to 0.9319 by round 3, indicating an increasing difficulty in detection. Furthermore, by augmenting the training set of the random forest classifier with these adversarial examples, the classifier’s efficacy in detecting DGA malware families not seen during training significantly improved, demonstrating the potential of adversarial training in enhancing malware detection models.

In \cite{botnet_2}, the authors designed CharBot, a novel domain generation algorithm (DGA) intended to bypass DGA classifiers effectively. CharBot operates by making simple, yet strategic modifications to benign domain names sourced from Alexa's top domain list. Its mechanism involves randomly altering two characters in these domain names to create new, unregistered domains that evade detection. This approach is significantly simpler than other DGAs, such as DeepDGA or DeceptionDGA, as it doesn't rely on deep learning or sophisticated string manipulations. Instead, CharBot's strength lies in its simplicity and the fact that it requires no knowledge of the targeted DGA classifiers, making it a fully black-box method. The authors tested CharBot's effectiveness against state-of-the-art DGA classifiers like FANCI (a random forest model) and LSTM.MI (a deep neural network model), demonstrates its capability to evade detection and highlight the inherent vulnerabilities in classifiers that only rely on domain name analysis. In their experimental evaluation of "CharBot: A Simple and Effective Method for Evading DGA Classifiers," the authors demonstrated that CharBot could effectively evade detection by advanced DGA classifiers. CharBot, when tested against state-of-the-art classifiers such as FANCI, and LSTM.MI, and B-RF, successfully bypassed their detection mechanisms. The classifiers, even after being retrained with CharBot-generated samples, failed to achieve practical detection rates against CharBot attacks. Specifically, the LSTM.MI model considered the most effective among the tested classifiers, could detect only around 55\% of CharBot samples at a false positive rate of 0.1\%, indicating a significant evasion capability by CharBot. Additionally, the study revealed that the simple string manipulation technique used by CharBot was more effective in evading detection compared to more complex methods used by other DGAs, highlighting a critical vulnerability in classifiers relying solely on domain name analysis. These results underscore the need for more robust detection systems that go beyond basic domain name string analysis to counter such simple yet effective adversarial attacks.

In \cite{botnet_3}, the authors designed an adversarial learning technique to evade Domain Generation Algorithm (DGA) classifiers. MaskDGA operates by adding perturbations at the character level to algorithmically generated domain names. The approach is based on training a substitute model on publicly available AGD datasets, then generating AGD names to which MaskDGA is applied. This involves constructing a Jacobian-based saliency map (JSM) by performing a single feed-forward and back-propagation step in the substitute model, identifying the characters with the highest gradient values in the JSM. MaskDGA then replaces exactly half of these characters in each domain name, resulting in new domain names that are falsely classified as benign by DGA classifiers. This method requires no prior knowledge of the classifier’s architecture and parameters, making it a practical, black-box technique for evading detection by security solutions at the network perimeter. In the experiment conducted for "MaskDGA: A Black-box Evasion Technique Against DGA Classifiers," the authors tested MaskDGA against four deep learning-based DGA classifiers and observed a substantial reduction in their detection efficacy. The average F1-score of these classifiers decreased from 0.977 to 0.495 when applying MaskDGA, indicating its high effectiveness in evading detection. Additionally, the study compared MaskDGA with two baseline attacks: a random attack and an attack based on DeepDGA. While the random attack demonstrated some level of effectiveness, MaskDGA significantly outperformed both, showcasing its superiority in generating domain names that evade detection. Furthermore, when evaluated against adversarial defenses like re-training and distillation, MaskDGA still proved to be a formidable challenge, suggesting that even enhanced classifiers struggle against this sophisticated evasion technique. These results highlight MaskDGA's effectiveness in deceiving advanced DGA classifiers and underscore the need for more resilient detection mechanisms.

In \cite{botnet_4}, the authors designed Khaos, a novel domain generation algorithm (DGA) based on neural language models and the Wasserstein Generative Adversarial Network (WGAN). Khaos consists of four major components: the n-gram Generator, n-gram Tokenizer, Domain Synthesizer, and Domain Filter. The n-gram Generator constructs a dictionary of common n-grams from real domain names, and the n-gram Tokenizer uses this dictionary to break down real domain names into n-grams. The Domain Synthesizer, operating on the WGAN framework, then learns to arrange these n-grams to create new domain names that mimic real domain structures. The final component, the Domain Filter, ensures the generated domain names are eligible for registration by discarding those that violate naming conventions or already exist. This innovative design allows Khaos to produce domain names with high anti-detection capabilities, effectively evading various state-of-the-art DGA detection methods. In the experimental evaluation of "Khaos: An Adversarial Neural Network DGA With High Anti-Detection Ability," the authors tested Khaos against several advanced DGA detection approaches and demonstrated its substantial ability to evade detection. Specifically, under the distribution-based detection approach, Khaos achieved an average detection distance of 0.64. Its performance against the statistics-based and LSTM-based detection approaches yielded AUCs (Area Under the Curve) of 0.76 and 0.57, respectively. Moreover, the precision of Khaos under the graph-based detection approach was recorded at 0.68. These results significantly surpassed the performance of other state-of-the-art DGAs, establishing Khaos's superiority in anti-detection capabilities. Additionally, the study revealed that training existing detection approaches with a dataset including domain names generated by Khaos could improve their detection ability, underscoring the potential of Khaos not only as a challenging DGA but also as a tool for enhancing the robustness of DGA detection methods.

In \cite{botnet_5}, the authors designed an evasion technique targeting deep learning models used for DGA detection. CLETer works by intelligently altering characters in existing DGA domain names to evade classification. It involves two primary steps: first, quantifying the influence of each character on the classification result using a scoring function, and second, applying character-level transformations to the most influential characters. This process effectively disguises malicious DGA domain names as benign, thereby evading detection. The scoring functions used include Head Influence Score (HIS), Tail Influence Score (TIS), Combined Influence Score (CIS), and Overall Influence Score (OIS). These functions are used to rank characters based on their impact on classification, guiding the choice of which characters to alter. CLETer's design emphasizes a black-box approach, where no internal knowledge of the targeted model is required, only the input-output relationship, making it broadly applicable and adaptable to different DGA classifiers.  In the experimental evaluation of "CLETer: A Character-level Evasion Technique Against Deep Learning DGA Classifiers," the authors demonstrated CLETer's potent ability to evade detection by deep learning DGA classifiers. When applying CLETer to known DGAs involved in the training phase, it notably reduced the recall of LSTM and CNN classifiers to extremely low levels. Specifically, with just two character modifications, CLETer lowered the LSTM classifier’s recall from 99.76\% to 1.29\% and the CNN classifier’s recall from 99.36\% to 3.64\%. Additionally, the study showed that CLETer was also effective against unknown DGAs, reducing recall significantly in these cases as well. The method proved to be more efficient than random substitution, emphasizing the effectiveness of strategic character modification. The research highlighted CLETer's capacity to mislead classifiers through minimal yet targeted alterations, underscoring a vulnerability in current DGA detection methodologies.

In \cite{botnet_6}, the researchers designed an experiment to analyze the transferability of adversarial attacks in computer networks, focusing on CNN-based models. They utilized three datasets: N-BaIoT, DGA, and RIPE Atlas, and two distinct network architectures, SPRITZ1 and SPRITZ2, to evaluate this transferability. The experiment involved launching five different adversarial attacks - I-FGSM, JSMA, LBFGS, PGD, and DeepFool - and examining their effectiveness across various scenarios, including cross-training, cross-model, and cross-model and training transferability. This approach allowed them to assess the impact of different network architectures and training datasets on the success of adversarial attacks. Additionally, the study explored defensive strategies, such as using mismatch classifiers and LSTM architectures, to understand how these factors influence the robustness of machine learning models against these attacks. In their experimental evaluation, the authors of "Demystifying the Transferability of Adversarial Attacks in Computer Networks" found notable outcomes regarding the transferability of adversarial attacks in computer networks. The experiments, using SPRITZ1 and SPRITZ2 architectures across N-BaIoT, DGA, and RIPE Atlas datasets, revealed that the transferability of adversarial attacks was significantly influenced by the mismatch in training datasets and network architectures. For instance, in the cross-model transferability scenario, attacks like JSMA and LBFGS demonstrated higher success rates when the same datasets were trained on different architectures, indicating effective transferability under certain conditions. However, in scenarios with complete mismatches in both networks and datasets, the transferability of attacks drastically reduced, suggesting limited effectiveness in more diverse settings. This analysis underscores the complexity of adversarial attack dynamics in CNN-based network models and highlights the importance of considering both architectural and dataset variances in assessing network security against such attacks.

\section{Adversarial examples in network intrusion detection}

DL-based Network Intrusion Detection Systems are threatened by adversarial attacks where attackers can disrupt and lead DL models to make incorrect decisions, given the extreme difficulty of detecting such adversarial manipulations. 
Such attacks inject imperceptible perturbations to the input that are enough to change the result of the prediction/classification. 
In the context of a black-box system, attackers employ a strategy of repeatedly altering small groups of traffic features without explicit knowledge of the targeted Network Intrusion Detection System (NIDS). They then query the system to generate adversarial samples. The feedback received after each query, such as an acknowledgment or lack thereof, serves as an indicator of the success or failure of the attack. This feedback is utilized to adapt the perturbations on specific traffic features, such as intervals between consecutive packets, or introduce new ones. The iterative process continues until the attacker successfully evades detection by the NIDS \cite{7jiang2022fgmd,7qiu2020adversarial,7shu2020generative}.
This approach allows malicious flows to camouflage themselves as benign traffic, enabling them to evade detection by NIDSs previously considered highly accurate. As a result, attackers can successfully compromise and disrupt their intended targets \cite{7ibitoye2019analyzing,7zhang2020tiki}.

In \cite{net_1}, the authors designed a comparative study to assess the resilience of deep learning models against adversarial attacks in IoT intrusion detection systems. They employed two neural network models: a traditional Feedforward Neural Network (FNN) and a Self-normalizing Neural Network (SNN) utilizing the Scaled Exponential Linear Unit (SELU) activation function. The study used the BoT-IoT dataset from UNSW Canberra Cyber's Cyber Range Lab, which simulates realistic IoT network traffic. To test the models' robustness, adversarial samples were crafted using methods like the Fast Gradient Sign Method (FGSM), Basic Iteration Method (BIM), and Projected Gradient Descent (PGD). The evaluation focused on various performance metrics, including accuracy, precision, recall, and multi-classification metrics like Cohen Cappa score, to analyze the impact of adversarial attacks on these models. Additionally, the study explored the effects of feature normalization on the adversarial resilience of the IDS models. This comprehensive design aimed to understand the vulnerability of deep learning-based intrusion detection systems in IoT environments to sophisticated adversarial attacks. In the study "Analyzing Adversarial Attacks Against Deep Learning for Intrusion Detection in IoT Networks," the researchers evaluated the performance of two deep learning-based Intrusion Detection Systems (IDS), a Feedforward Neural Network (FNN) and a Self-normalizing Neural Network (SNN), against adversarial attacks using the BoT-IoT dataset. The FNN initially achieved a 95.1\% accuracy but saw a significant decrease in performance when subjected to adversarial attacks, with accuracies dropping to 24\%, 18\%, and 31\% under FGSM, BIM, and PGD attacks, respectively. While the FNN outperformed the SNN in standard classification metrics, the SNN displayed greater resilience to adversarial samples, with an average of 9\% higher performance accuracy against such attacks. The study also revealed that feature normalization, while improving performance metrics, increased the vulnerability of both IDS models to adversarial samples, underscoring the trade-offs in IDS design for IoT networks.

In \cite{net_2}, the researchers designed a comprehensive study to evaluate and improve the robustness of deep learning-based Network Intrusion Detection Systems (NIDS) against adversarial attacks. Utilizing the CSE-CIC-IDS2018 dataset, the study involved testing three neural network architectures – Multilayer Perceptron (MLP), Convolutional Neural Network (CNN), and Long Short-Term Memory (LSTM) – against five state-of-the-art adversarial attack methods, including Natural Evolution Strategies (NES), Boundary Attack, HopSkipJumpAttack, Pointwise, and Opt-Attack. These attacks were conducted in both one-to-all and one-to-one detection scenarios. To counter these attacks, the study introduced three defense mechanisms: model voting ensembling, ensembling adversarial training (EAT), and query detection, assessing their effectiveness using metrics like Attack Success Rate (ASR), Mean Absolute Percentage Error (MAPE), and average number of queries. The research also considered domain constraints to ensure the practicality and integrity of the adversarial samples. In the study "Tiki-Taka: Attacking and Defending Deep Learning-based Intrusion Detection Systems," extensive testing revealed that adversarial attacks could successfully evade Network Intrusion Detection Systems (NIDS) with attack success rates (ASR) as high as 35.7\% by merely altering time-based traffic features. After implementing Ensemble Adversarial Training (EAT), the detection performance of NIDS models slightly dropped in accuracy, precision, and F1 score, but their recall rate improved, indicating an increased tendency to classify ambiguous samples as anomalies. This led to higher false positive rates but lower false negatives. The defense mechanisms proposed, including model voting ensembling, EAT, and query detection, significantly enhanced intrusion detection, as evidenced by the substantial reduction in ASR across various attacks. These defenses also improved the models' resilience to older adversarial samples, showcasing the effectiveness of EAT in fortifying NIDS against such sophisticated attacks.

In \cite{net_3}, the authors designed the Generative Adversarial Active Learning (Gen-AAL) method to create adversarial attacks against black-box Intrusion Detection Systems (IDS). This method combined a Generative Adversarial Network (GAN) with an active learning algorithm. The GAN consisted of a Variational Autoencoder (VAE) for generating adversarial feature points and a Multi-Layer Perceptron (MLP) model simulating the black-box IDS, referred to as the Substitute-IDS (S-IDS). The active learning component of Gen-AAL was used to iteratively retrain the VAE and S-IDS models using a limited number of labeled data points queried from the actual IDS. This design aimed to efficiently generate adversarial feature points that could bypass the IDS while minimizing the required labeled data and not needing prior knowledge of the IDS model's internal structure or loss function. The study's approach focused on balancing the training of the VAE model and the S-IDS model through hyperparameter tuning to achieve high success rates in adversarial attacks. In their study "Generative Adversarial Attacks Against Intrusion Detection Systems Using Active Learning," the researchers demonstrated the effectiveness of the Gen-AAL algorithm in bypassing black-box Intrusion Detection Systems (IDS). The Gen-AAL algorithm achieved a significant 98.86\% success rate in evading the IDS model with only one iteration of model retraining and using just 25 labeled data points. This success rate was notably higher than the DeepFool Adversarial Learning (DFAL) algorithm, which achieved a 95.48\% success rate at its first iteration, increasing to 96.25\% at the third iteration with 45 labels used. The Gen-AAL algorithm also exhibited a higher average l2-distance perturbation of 1.174, compared to DFAL’s 0.506. Additionally, the study highlighted the sensitivity of the adversarial attack success rate to hyperparameters related to model training, emphasizing the importance of fine-tuning these parameters for optimal performance.

In \cite{net_4}, Han Qiu and colleagues developed a two-phase method to conduct adversarial attacks against deep learning-based Network Intrusion Detection Systems (NIDS) in IoT environments. Initially, they used a model extraction technique to replicate the black-box NIDS model with minimal training data, significantly enhancing the attack's efficiency. This phase involved reconstructing the Feature Extraction (FE) module and collecting network traffic packets to train a new shadow model. In the second phase, they generated adversarial examples (AEs) using saliency maps to pinpoint critical features influencing the NIDS detection results. These features were then strategically perturbed using methods like the Fast Gradient Sign Method (FGSM), creating AEs that could effectively bypass the NIDS while requiring minimal modifications to the network packets. This approach, tailored for IoT contexts, underscores the vulnerabilities of deep learning-based NIDS to sophisticated adversarial attacks. In "Adversarial Attacks Against Network Intrusion Detection in IoT Systems," Han Qiu and colleagues demonstrated a successful adversarial attack against the Kitsune Network Intrusion Detection System (NIDS). Their method, requiring less than 0.005\% modification in malicious packet bytes, achieved an impressive 94.31\% attack success rate. The efficiency of their approach was further highlighted as they replicated the NIDS's shadow models using only 10\% of the original training data. The average computation time for generating adversarial examples was minimal, indicating the attack's practicality for real-time scenarios. For example, in DoS attack scenarios, the adversarial example generation caused only a 0.171-second delay. They also found that increasing the packet dropping rate led to a significant rise in false positives, with higher rates causing more normal packets to be misclassified as malicious. This study reveals critical vulnerabilities in deep learning-based NIDS, particularly in IoT environments, and underscores the need for developing more resilient defense mechanisms.

\section{Adversarial examples in user identification and authentication}

More recent studies \cite{8zhang2021attack,8zhang2020voiceprint,8tan2019adversarial,8yin2021adv} have shown that user identification/authentication systems are also vulnerable to adversarial attacks. 
\cite{8szegedy2013intriguing} showed that tiny (almost invisible) perturbations on images could be crafted to mislead a state-of-the-art DNN-based classifier.

The design of the study in \cite{user_1} by Yi Xiang Marcus Tan et al. involves evaluating the vulnerability of mouse dynamics-based user authentication systems to adversarial attacks. The authors employ three distinct attack strategies: statistics-based, imitation-based, and surrogate-based. They use two comprehensive mouse dynamics datasets, Balabit and The Wolf of SUTD (TWOS), for their experiments. For the imitation-based attacks, a Gated Recurrent Unit (GRU) variant of Recurrent Neural Networks (RNNs) is utilized to generate adversarial mouse movement sequences. The surrogate-based attacks involve training a surrogate classifier, distinct from the target model, to generate adversarial samples. The experiments are conducted in a realistic scenario where the attacker has access to user data but not to the authentication model or its outputs. This design allows for a thorough examination of the robustness of mouse dynamics authentication models against different types of adversarial strategies. In the study "Adversarial Attacks on Remote User Authentication Using Behavioural Mouse Dynamics," the authors found that all three attack strategies—statistics-based, imitation-based, and surrogate-based—were effective to varying degrees in compromising the mouse dynamics authentication models. The statistics-based attack achieved a notable success rate, with at least 26.4\% of adversarial samples bypassing the target classifiers, and up to 61.54\% in some cases. The imitation-based attack, leveraging a Gated Recurrent Unit (GRU) model, demonstrated improved effectiveness over the statistics-based approach, particularly for longer sequence lengths. This suggests that the user's distinctive traits in mouse dynamics are more prominent over extended time frames. The surrogate-based attacks, using both GRU-RNN and Fully Connected (FC) surrogate models, performed better than the statistics-based method in bypassing target classifiers. However, the success rate varied depending on the dataset and the classifier type. These results highlight the vulnerability of mouse dynamics-based authentication systems to adversarial attacks, emphasizing the need for more robust security measures in behavioral biometric systems.

In the study \cite{user_2}, Lei Zhang et al. designed an experiment to test the effectiveness of VMask, a novel voiceprint mimicry attack strategy. The design focused on evaluating VMask's ability to bypass automatic speaker verification (ASV) systems used in smart homes. The experiment was structured in two parts: a grey box attack on VGGVox, a known ASV system, and a black box attack targeting Microsoft Azure Speaker Verification (MS-ASV). Additionally, a real-world case study was conducted on the Apple HomeKit platform to assess VMask's practical applicability. The grey box attack utilized Zeroth Order Optimization (ZOO) for generating adversarial audio, aiming to achieve high confidence scores while preserving the original content of the audio. For the black box attack, the team crafted adversarial audios that mimicked different speakers' voiceprints to deceive the ASV system, demonstrating VMask's effectiveness in a scenario where the internal workings of the target system are unknown. This comprehensive design allowed for a thorough evaluation of VMask's capabilities in both controlled and real-world environments. The experimental outcomes of "Voiceprint Mimicry Attack Towards Speaker Verification System in Smart Home" by Lei Zhang et al. demonstrated the effectiveness of VMask, their proposed voiceprint mimicry attack. In the grey box attack against VGGVox, VMask achieved a 95\% success rate. The adversarial manipulations applied to the audio samples substantially increased the confidence scores by an average of 306\%, while maintaining an average Signal-to-Noise Ratio (SNR) of 13.13dB. This indicated that the original audio content was preserved effectively. In the black box attack on Microsoft Azure Speaker Verification (MS-ASV), VMask attained a 70\% success rate at the phrase level. During this attack, adversarial audios crafted from one speaker to mimic other speakers successfully bypassed the MS-ASV API in a majority of the trials. Furthermore, in a real-world case study involving Apple HomeKit, VMask demonstrated a 67.5\% average success rate across various Apple devices, underlining its practical applicability in smart home environments. These results collectively highlight VMask's potential as a robust tool for bypassing speaker verification systems, posing significant security implications for smart home devices.

In their study \cite{user_3}, Weiyi Zhang et al. designed a novel approach involving the creation of universal adversarial perturbations. Their method involved a two-step algorithm optimized for these perturbations to be text-independent and minimally affect speech content recognition. The design aimed to ensure that the perturbations could be played as a separate audio source during an attack, thereby bypassing audio replay detection systems commonly used in practical speaker verification (PSV) scenarios. The first step focused on maximizing the attack's impact on the speaker verification model, using a variety of audios from the adversary to ensure the perturbation’s effectiveness across different speech contents. In the second step, they optimized the perturbation to reduce its impact on speech recognition, ensuring the perturbation remains effective without significantly altering the audio content. This comprehensive approach aimed to craft perturbations that are not only universal (effective regardless of the speech content) but also robust (effective after being played over the air), thus addressing the key challenges in attacking PSV systems in real-world scenarios. The experimental outcomes of the study "Attack on Practical Speaker Verification System using Universal Adversarial Perturbations" by Weiyi Zhang et al. demonstrated the effectiveness of their two-step algorithm for generating adversarial perturbations against practical speaker verification systems. In digital attacks, the adversarial perturbations achieved high success rates, with intra-gender attacks showing a 98.43\% success rate and inter-gender attacks a 96.63\% success rate, while maintaining relatively low word error rates (WER) of 19.43\% and 21.53\%, respectively. These results indicate that the perturbations were effective in misleading the speaker verification system without significantly impacting speech content recognition. In physical attack scenarios with volunteers, the adversarial perturbations achieved a 100\% success rate in targeted attacks. Notably, these attacks could pass audio replay detection, and the increase in WER was only 3.55\% compared to clean speech, showcasing the perturbations' effectiveness in a live environment. The study thus highlighted the vulnerability of practical speaker verification systems to sophisticated adversarial attacks while maintaining the intelligibility and recognition of the speech content.

In their research on Adv-Makeup \cite{user_4}, Bangjie Yin et al. designed an innovative method to create imperceptible and transferable adversarial attacks on face recognition systems. The core design involved generating eye shadow-like perturbations in the orbital region of facial images. This was achieved through a makeup generation module using Generative Adversarial Networks (GANs) to synthesize realistic eye shadow, and a makeup blending strategy to ensure the natural integration of this synthetic makeup with the source face. To address the challenge of transferability, particularly in black-box scenarios, the team employed a fine-grained meta-learning adversarial attack strategy. This strategy enabled the learning of generalized adversarial features across various models, thereby enhancing the effectiveness of the attacks against different face recognition systems. The design aimed to create adversarial examples that were both visually imperceptible and highly effective in fooling advanced face recognition technologies under both digital and physical conditions. The experimental outcomes of "Adv-Makeup: A New Imperceptible and Transferable Attack on Face Recognition" by Bangjie Yin et al. demonstrated the efficacy of the Adv-Makeup technique in generating adversarial attacks on face recognition systems. The study reported high success rates in both digital and physical attack scenarios. In digital attacks, Adv-Makeup outperformed several existing attack methods, including FGSM, PGD, MI-FGSM, C\&W, and face-specific attacks like Adv-Glasses and Adv-Hat, in terms of attack success rate (ASR) across multiple models (IR152, IRSE50, MobileFace, FaceNet). Notably, the ASR of Adv-Makeup ranged from 7.59\% to 63.74\% across different datasets and target models, significantly higher compared to other methods. In physical attack scenarios, Adv-Makeup was tested against commercial face recognition platforms (Face++ and Microsoft), where it achieved substantial success. The attacks were robust to various real-world factors such as illumination, pose, gender, and makeup intensity. The study concluded that Adv-Makeup could generate adversarial examples that were both imperceptible and highly transferable, effectively fooling advanced face recognition systems in practical settings.

In the study \cite{user_5} by Haibin Wu et al., the researchers designed a novel detection framework for identifying adversarial samples in automatic speaker verification (ASV) systems. This framework employs neural vocoders, specifically Parallel WaveGAN, to re-synthesize audio from acoustic features. The detection process involves comparing the ASV scores of the original audio and the re-synthesized audio. A significant discrepancy between these scores is used to flag adversarial samples. The study explores the effectiveness of this approach across various attack intensities, represented by different values of \(\epsilon\). The neural vocoder's ability to model genuine data distributions and reduce distortion in re-synthesized audio plays a crucial role in this detection strategy. This method stands out for not requiring prior knowledge of the adversarial attack methods, thereby offering a more practical solution in real-world scenarios. The researchers also compare their approach with traditional methods, like the Griffin-Lim algorithm, and alternative attack-agnostic methods like Gaussian filtering, to establish the superiority of their neural vocoder-based framework in detecting adversarial attacks on ASV systems. The experimental outcomes of "Adversarial Sample Detection for Speaker Verification by Neural Vocoders" demonstrated the effectiveness of using neural vocoders for detecting adversarial samples in automatic speaker verification (ASV) systems. The study utilized Parallel WaveGAN as a neural vocoder and compared it with traditional methods like the Griffin-Lim algorithm. The results showed that the neural vocoder-based method significantly outperformed the Griffin-Lim algorithm in detecting adversarial samples across various attack intensities (\(\epsilon\)). For example, with an attack intensity of \(\epsilon = 20\), the detection rate using the neural vocoder was as high as 99.76\%, indicating a high level of effectiveness. Furthermore, the study established that the neural vocoder method was robust and effective under different attack scenarios, with consistently high detection rates even at lower attack intensities (\(\epsilon = 5\)). This superiority of the neural vocoder-based detection method was attributed to its ability to model the genuine data distribution and reduce distortion in the re-synthesized audio, leading to a more accurate distinction between genuine and adversarial samples. The research also demonstrated that the neural vocoder approach is dataset-independent, enhancing its applicability in diverse ASV systems.

\section{Adversarial examples in encrypted traffic analysis}

Due to the effectiveness of deep learning technology in encrypted traffic analysis and the unique characteristics of encrypted traffic itself, attackers may employ clever tactics to craft adversarial examples, making it challenging for deep learning models to accurately classify and analyze encrypted communications. This situation raises concerns about the robustness of deep learning models against adversarial examples, particularly when maintaining high levels of accuracy and reliability is crucial in dealing with encrypted traffic. Some recent work mentioned this problem \cite{9rahman2020mockingbird,9usama2019black,9verma2018network}.

In their study on network traffic obfuscation using adversarial machine learning \cite{encry_1}, Verma et al. designed an experiment to evaluate the effectiveness of adversarial perturbations in misleading traffic classifiers. The core of the experiment involved using a neural network classifier trained on a dataset comprising various network traffic types, including packet sizes and inter-packet arrival times. The authors employed the Carlini-Wagner L2 algorithm, an adversarial machine learning technique, to generate modified traffic samples. These samples were crafted to minimally deviate from the original traffic in terms of statistical features, thereby ensuring minimal disruption to network performance and adherence to protocol constraints. The goal was to create obfuscated traffic (T') that would be classified differently from the original traffic (T), without being easily discernible from T in terms of packet size and time distributions. This experimental design allowed the researchers to systematically evaluate the feasibility and efficiency of using adversarial machine learning to obfuscate network traffic, thus protecting it from classification by potential attackers. The experimental outcomes of the study "Network Traffic Obfuscation: An Adversarial Machine Learning Approach" by Gunjan Verma et al. demonstrated the efficacy of adversarial machine learning techniques in obfuscating network traffic. The study employed the Carlini-Wagner L2 algorithm to create adversarial examples of network traffic, which were then tested against a neural network classifier trained on a variety of network traffic types. The results showed high success rates in misleading the classifier, with a significant proportion of obfuscated traffic being misclassified as different types from the original. For instance, a considerable fraction of obfuscated database traffic was classified as WWW traffic, indicating the effectiveness of the obfuscation. The perturbations introduced by the adversarial examples were minimal, ensuring that the obfuscated traffic (T') remained statistically similar to the original traffic (T), particularly in terms of packet size distribution. The inter-arrival times, although slightly more altered, were still within a reasonable range of the original traffic. These findings underscored the potential of adversarial machine learning in creating effective yet minimally intrusive obfuscations for network traffic, offering a promising approach to enhancing network security and privacy.

In their study on black-box adversarial machine learning attacks on network traffic classification \cite{encry_2}, Muhammad Usama et al. designed a method to assess the robustness of ML/DL techniques used in network traffic classification under adversarial conditions. The experiment was conducted using the Tor-NonTor dataset for both binary and multi-class traffic classification. They employed Deep Neural Networks (DNNs) and Support Vector Machines (SVMs) for the classification tasks. The adversarial attack methodology involved crafting adversarial samples using mutual information for feature selection. The crafted samples were designed to minimally but effectively alter the traffic samples, aiming to mislead the classifiers without any prior knowledge of the ML model's training data or architecture (black-box approach). The goal was to demonstrate the susceptibility of ML/DL-based traffic classification systems to adversarial perturbations, thus highlighting potential security risks in network applications relying on these techniques. The experimental outcomes of "Black-box Adversarial Machine Learning Attack on Network Traffic Classification" by Muhammad Usama et al. revealed significant vulnerabilities in ML/DL-based network traffic classification systems when exposed to black-box adversarial attacks. Using the Tor-NonTor dataset, the study tested both binary (Tor vs. Non-Tor) and multi-class traffic classification scenarios using Deep Neural Networks (DNNs) and Support Vector Machines (SVMs). The results showed a marked decrease in classification accuracy post-attack: for the DNN-based binary classifier, accuracy dropped from 96\% to 77\%, and for the SVM-based classifier, from 93.54\% to 77.41\%. In the multi-class classification scenario targeting the 'Chat' class, accuracy for the DNN-based classifier plummeted from 96.3\% to 2\%, and for the SVM-based classifier, from 96.4\% to 63.95\%. These findings highlight the effectiveness of adversarial ML attacks in significantly compromising the integrity of network traffic classification systems, underscoring the urgent need for more robust defense mechanisms against such threats in ML/DL-based network applications.

In the study \cite{encry_3}, the researchers developed a defense mechanism against Website Fingerprinting (WF) attacks on systems like Tor. The design of Mockingbird involves generating adversarial network traffic traces that resist classification by deep learning classifiers used in WF attacks. This approach increases the randomness in the generation of these traces, deliberately moving in unpredictable patterns in the space of viable traffic traces. Unlike traditional methods that follow predictable gradients, Mockingbird's design avoids this, making it harder for an attacker to adapt their classifier through adversarial training. The adversarial traces are created to have high misclassification rates with small perturbations, ensuring low overhead. The research aimed to demonstrate that Mockingbird could significantly reduce the accuracy of state-of-the-art WF attacks, even those hardened with adversarial training, while maintaining reasonable bandwidth overhead, thereby offering an improved defense over existing methods like WTF-PAD and Walkie-Talkie. The experimental outcomes of "Mockingbird: Defending Against Deep-Learning-Based Website Fingerprinting Attacks With Adversarial Traces" demonstrated the effectiveness of the Mockingbird defense against WF attacks. The technique significantly lowered the accuracy of deep learning classifiers used in WF attacks, hardened with adversarial training. Specifically, it reduced the classification accuracy from 98\% to a range of 38-58\%, indicating a substantial decrease in the classifier's ability to accurately identify websites visited over Tor. Additionally, this was achieved with a bandwidth overhead of only 58\%, which is a marked improvement over existing defense mechanisms like WTF-PAD and Walkie-Talkie. The results underscored Mockingbird's capability to generate adversarial traces that effectively confuse classifiers while maintaining a balance between defense robustness and network efficiency. This balance is critical in real-world applications, where excessive bandwidth overhead can be as detrimental as weak defense mechanisms.

In their research on defending against Website Fingerprinting (WF) attacks \cite{encry_4}, Chengshang Hou and colleagues designed the Attack2Attack (A2A) framework, which utilizes adversarial examples to confuse the attacker's classifier. A2A comprises two main components: a traffic distorter and a traffic detector. The traffic distorter operates by inserting dummy packets into the traffic sequence at random positions. This is done iteratively, with the distorter adjusting its strategy based on feedback from the traffic detector. The detector itself employs a substitute model that mimics the classification boundary of the attacker's model. This model is used to classify the distorted traffic and provide feedback to the distorter for improving the adversarial examples. The design of A2A focuses on crafting adversarial examples that are effective in a black-box setting, where the defender lacks knowledge about the attacker's model specifics. The framework aims to reduce the accuracy of the attacker's classifier while incurring minimal bandwidth overhead, thus providing an efficient and robust defense against WF attacks. The experimental outcomes of "Attack versus Attack: Toward Adversarial Example Defend Website Fingerprinting Attack" demonstrated the effectiveness of the Attack2Attack (A2A) defense framework. A2A significantly reduced the accuracy of the attacker's model in website fingerprinting attacks. Specifically, using the best substitute model, the attacker's accuracy was reduced to as low as 57.6\% with only a 2.2\% increase in bandwidth overhead. This performance significantly surpasses manually designed defenses which typically have much higher bandwidth overheads (31\% on average). Additionally, the study explored the transferability of adversarial examples among various models. It found that adversarial examples generated with the DF-adamax model as a substitute were particularly effective, reducing the accuracy of other models to varying degrees. The experiments also compared the performance of A2A under different settings (bidirectional and unidirectional, ordered and random candidate sets), concluding that bidirectional ordered vector settings achieved the best transferability, leading to an average accuracy decline to 57.6\% in other models. These results highlight A2A's effectiveness in reducing classification accuracy with minimal bandwidth costs, outperforming existing WF defense mechanisms.

In \cite{encry_5}, Sadeghzadeh, Shiravi, and Jalili designed an experimental approach to test the robustness of DL-based network traffic classifiers against adversarial attacks. The study introduces three novel attacks: AdvPad, AdvPay, and AdvBurst, each targeting different aspects of network traffic classification—packet classification, flow content classification, and flow time series classification, respectively. The AdvPad attack injects a Universal Adversarial Perturbation (UAP) into packet contents, AdvPay into the payload of a dummy packet, and AdvBurst into a series of dummy packets within a flow's burst. These attacks were developed to evaluate how small perturbations in network traffic could significantly degrade the performance of DL classifiers. The researchers' methodology focused on creating adversarial examples that would be realistic in a network environment and could be generated live, without the need for buffering the entire network traffic. This approach allowed for a comprehensive assessment of the vulnerabilities present in current DL-based network traffic classifiers. The experimental results from "Adversarial Network Traffic: Towards Evaluating the Robustness of Deep-Learning-Based Network Traffic Classification" demonstrated the vulnerability of DL-based network traffic classifiers to adversarial attacks. The three proposed attacks—AdvPad, AdvPay, and AdvBurst—effectively decreased the classifiers' performance. The AdvPad attack, which injected UAP into packet contents, significantly reduced classifier accuracy. Similarly, the AdvPay attack, targeting the payload of dummy packets, also led to a notable decrease in classification accuracy. The most impactful was the AdvBurst attack, which involved inserting dummy packets with crafted statistical features into a flow's burst; this attack drastically lowered the accuracy of classifiers analyzing flow time series. These results highlighted that even minimal adversarial perturbations could significantly compromise the effectiveness of DL-based network traffic classifiers across various classification categories. The study underscored the need for enhanced robustness in network traffic classifiers against such adversarial threats.

\section{Countermeasures}

In general, defense mechanisms are designed based on one of the following two main approaches: (i) Passive defense: deploying defensive strategies after an attack occurs (e.g., forgetting data observed during training after a poisoning attack), and (ii) Active defense: strengthening system defenses against potential threats before an actual attack takes place. Most defense systems follow the latter approach to minimize damage. Therefore, we summarize some defense techniques.

Modify Classifier: The gradient masking approach involves altering the ML model to make its gradients less susceptible to attacks from adversaries. It has been demonstrated that applying gradient masking to saturated S-shaped networks may lead to the disappearance of gradient effects \cite{duddu2018survey}. Defense distillation techniques \cite{hinton2015distilling} enable the transfer of knowledge from larger neural networks to smaller ones. The output of the original neural network can be utilized to fine-tune this technique, training a smaller network to resist adversarial examples. Given the success of this method in the field of computer vision, it is essential to conduct a thorough assessment of its applicability in security domains such as malicious software detection, particularly in the face of adversarial examples.

Introducing a dedicated detection approach: \cite{xu2017feature} introduced an innovative strategy known as feature compression to effectively counter adversarial examples. This technique revolves around the fundamental concept of compressing the features of a given sample and subsequently subjecting the compressed sample to classification. The key criterion for identifying adversarial examples lies in the substantial disparity between the classifier's output for the compressed sample and that for the original sample. Any notable difference in these outputs designates the latter as an adversarial example. This method represents a proactive measure to enhance the robustness of classifiers against potential adversarial attacks.

\section{Conclusion}

Our investigation demonstrates that ML/DL applications play a significant role in enhancing cybersecurity, particularly in data-driven learning and the execution of complex real-time tasks. A literature review underscores the effectiveness of attacks utilizing adversarial examples against ML/DL-based security systems, resulting in potential performance deterioration. This paper delves into the influence of adversarial examples on the fundamental aspects of cybersecurity applications grounded in deep learning. With a specific focus on cybersecurity, our objective is to meticulously scrutinize the present state of such attacks and provide support to the research community by highlighting pertinent scientific achievements. Our overarching goal is to develop and deploy intelligent systems that efficiently harness vast amounts of generated data, ensuring the stability, resilience, and security of underlying ML/DL models. This approach takes into account the unique characteristics of network security. To achieve this, we scrutinize adversarial attack types, methods for generating adversarial examples, and corresponding defense strategies, summarizing the latest research. We introduce a classification system for various network security applications, reviewing recent studies on adversarial example-based attacks and effective defenses against them. This comprehensive approach aims to advance and facilitate the secure adoption of promising ML/DL solutions in cybersecurity.

\vspace{12pt}

\end{document}